\documentclass[twocolumn]{aastex631}

\begin{document}

\title{Impact of Starlink constellation on Early LSST: a Photometric Analysis of Satellite Trails with BRDF Model}

\correspondingauthor{Yao Lu}
\email{luyao@pmo.ac.cn}

\author[0000-0003-0352-3790]{Yao Lu}
\affiliation{Purple Mountain Observatory, Chinese Academy of Sciences, Nanjing 210023, China}
\affiliation{Key Laboratory for Space Objects and Debris Observation, Chinese Academy of Sciences,
Nanjing 210023, China}



\begin{abstract}
We report a simulation and quantification of the impact of the Starlink constellation on LSST in terms of the trail surface brightness using a BRDF-based satellite photometric model. A total of 11,908 satellites from the Gen1 and Gen2A constellations are used to focus on the interference to the initial phase of LSST operation. The all-sky simulation shows that approximately 69.33\% of the visible satellites over station have an apparent brightness greater than 7 mag with a v1.5 satellite model. The impact of satellite streaks exhibit a non-monotonic relationship to the solar altitude, with the worst moments occurring around $-15^{\circ}$ solar altitude. The assessment based on simulated schedules indicates that no trails can reach the saturation-level magnitude, but 71.61\% trails show a surface brightness brighter than the best-case crosstalk correctable limits, and this percentage increases as the dodging weight increases. Therefore, avoiding satellites in the scheduler algorithm is an effective mitigation method, but both the number of streaks and their brightness should be taken into account simultaneously.
\end{abstract}



\section{Introduction} \label{sec:intro}

The impact of mega Low-Earth-Orbit (LEO) constellations on ground-based optical astronomy depends on the number of visible satellites and their brightness \citep{darkandquietskies1,darkandquietskies2,Green2022}. Quantitative assessment can verify the efficacy of mitigation techniques and lead to further strategies in collaboration with satellite operators \citep{satcon1,satcon2}. It can also help astronomers address the loss of science and explore possible mitigation measures \citep{Tyson2020a,Mroz2022b,Kruk2023,Hainaut2020,McDowell2020}, such as adding avoidance of bright satellites to the schedule \citep{Hu2022}. 

The upcoming Legacy Survey of Space and Time (LSST) by the Rubin Observatory will be significantly impacted, particularly for the near-Earth object (NEO) searches, which is a major scientific objective and will be conducted under highly visible satellite conditions. While a correction algorithm can largely eliminate the nonlinear crosstalk caused by satellite trails, it will be ineffective for over-bright trails \citep{Tyson2020a}. Quantitative modelling of the surface brightness of trails appearing on LSST images is therefore crucial for the assessment. The effect of Starlink constellation on LSST is of great interest due to it being the largest internet constellation currently in orbit and still rapidly deploying. This work adopts a more reliable short-term launch plan given the frequent adjustments to constellation structures. The plan includes the entire 1st Generation (Gen1) and Phase 1 of 2nd Generation (Gen2A), totalling 11,908 satellites \citep{McDowell_web_2024}. In other words, we focus primarily on the impact on the earliest years of the LSST program.

Predicting the transient magnitude of a satellite at the moment of observation is usually difficult due to various factors affecting its brightness. To improve accuracy, astronomers are encouraged to make multiple observations from different angles and develop precise photometric models \citep{satcon1,satcon2}. These models include diffuse spherical models \citep{Lawler2022b,Osborn2022}, phase polynomial models \citep{mallama2021}, Minnaert models \citep{Horiuchi2020,Tregloan-Reed2021a}, and shape-based BRDF models \citep{Cole,Fankhauser2023}. This work utilises a BRDF model from \citet{lu2024} to enhance the accuracy of satellite brightness under various observing scenarios. 

Surface brightnesses of trails, rather than satellite apparent magnitude (e.g. 7 mag for 550km satellite), can more accurately represent the different levels of impacts because they can directly convert to the ADU counts of the detector. Thus, we use the surface brightness thresholds obtained from \citet{Tyson2020a} to categorize streaks with different impact levels in both all-sky observations and simulated LSST schedules. This analysis will help to further comprehend the impact of Starlink satellites on LSST observations.

In this letter, the methodology for calculating the peak surface brightness of a trail and the brightness thresholds for LSST filters is described in Sec~\ref{sec:method}. A statistics on the surface brightness of streaks in LSST observations is presented in Sec ~\ref{sec:result}, including the proportion and distribution of overbright streaks.

\section{methodology} \label{sec:method}
\subsection{Number of satellite trails}\label{subsec:density}
An analytical method described in \citet{Bassa2022} is used to determine the number of satellite trails in observations. This quantity depends on the satellite density, the size of the field of view (FOV), the apparent angular motion of the satellite, and the exposure time. The density of satellites in a constellation shell, as observed from a particular direction, is determined by the total number of satellites, their orbital inclination and altitude, the latitude of the station, and the impact angle of the observation direction, as formulated in the appendix of \citet{Bassa2022}. Table~\ref{tab:orbits} lists the short-term plan for the Starlink constellation used in this letter, which contains 11,908 satellites distributed in $8$ shells. The total number of trails in an observation is calculated by adding up different shells. 
\begin{deluxetable}{lccccc}								
\tablecaption{Short-term planning for the Starlink Constellation. The parameters are collected from \citet{McDowell_web_2024}. It contains a total of 11,908 satellites, with the largest number of launched satellites being v1.5 type, as of 12 January 2024.\label{tab:orbits}}			
\tablewidth{0pt}								
\tablehead{								
\colhead{Phase} &\colhead{Group}  & \colhead{$N_{\rm Sat}$} & \colhead{Incl [$^{\circ}$]} & \colhead{Alt [km]} & \colhead{Sat Type}}								
\startdata								
Gen1&	G1      & 1584 	&	53    &	550     & v1.0	\\	
	&	G2      & 720 	&	70    &	570     & v1.5	\\	
	&	G3      & 348 	&	97.6  &	560     & v1.5	\\	
	&	G4      & 1584 	&	53.2  &	540     & v1.5	\\	
	&	Unknown & 172 	&	97.6  &	560     & Unknown\\ \hline
Gen2A&	G5\tablenotemark{a}   & 2500  &	43    &	530 & v1.5\&v2.0mini \\	
	&	G7      & 2500	&	53    &	525     & v2.0mini\\	
	&	G8      & 2500	&	33    &	535     & Unknown\\	
\enddata		
\tablenotetext{a}{Here group 5 includes group 6, totalling 2,500 satellites, all assumed at 530km orbits.}
\end{deluxetable}						

\subsection{Apparent magnitude of satellites}\label{subsec:apparent}
Starlink Gen1 has been completed with v1.0 and v1.5 satellites. Gen2A includes hundreds of v1.5 satellites, but v2.0-mini (or v2.0) will complete the remaining deployments and replace the defunct satellites. The pace of replacement, however, depends on the lifetimes of earlier satellites and the launch capabilities of new rockets. The v1.5 satellite is used throughout this work to predict satellite brightness. This choice is based not only on its largest number in orbit compared to other versions (as of January 2024), but also on its moderate brightness, which helps to reduce statistical bias in impact evaluation.  

High-fidelity satellite models, such as those based on BRDF, can often yield more accurate photometric predictions \citep{Cole,Fankhauser2023}. In this study, we use the photometric model of v1.5 satellite from \citet{lu2024}, which takes into account the chassis shading on the solar array and the earthshine effect. Assuming all satellites are in normal operation, angular arguments for the BRDF model can be derived from satellite's position and attitude \citep{lu2024}. The prototype magnitude given by the model is corrected for distance and atmospheric extinction to obtain the apparent magnitude of satellite. This model is based on the MMT satellite photometric data \citep{karpov2016}, which closely approximates the magnitude in the Johnson-V filter \citep{mallama2021}. To cover all filters used in LSST, the satellite magnitude in these filters is estimated from its Johnson-V magnitude. Multi-color observations from \citet{Mroz2022b} indicate that the colors of Starlink satellites are generally consistent with those of the Sun. Assuming that the satellite reflectance spectrum is essentially unchanged from the solar spectrum, we can convert the satellite magnitude by the solar AB magnitude in these filters \citep{Willmer2018}.

\subsection{Peak surface brightness of satellite trails}\label{subsec:surface}

The surface brightness of the satellite as it appears on the detector will be affected by the satellite's angular motion and the out-of-focus effect \citep{Ragazzoni2020b}. The effective magnitude of the trails caused by apparent angular motion is given by:
\[
m_{\rm eff} = m_{\rm app}-2.5\lg10(\frac{t_{\rm eff}}{t_{\rm exp}})
\]
where $m_{\rm app}$ is apparent magnitude and $t_{\rm exp}$ is exposure time, $t_{\rm eff}$ is the effective time that the blurred satellite (same as width of trail) crosses a pixel. Defocusing causes the width of the trail to extend as \citep{Tyson2020a}
\[
\theta_{\rm{sat}}^2 = {\rm{seeing}^2} + \frac{D_{\rm sat}^2+D_{\rm mir}^2}{\rho_{\rm sat}^2}
\]
and the effective time for the center of the trail is thus given by $t_{\rm eff} = \theta_{\rm sat}/\omega_{\rm sat}$, where $D_{\rm sat}$ is satellite projection size (set to $2$m), $\rho_{\rm sat}$ the range to satellite, $\omega_{\rm sat}$ the velocity of the satellite in the focal plane. The typical width of a trail on LSST detector is several arcsec for 550km satellites, i.e. tens of times larger than the pixel size. Assuming a signal spread area of $\pi(\theta_{\rm sat}/2)^2$ due to the defocusing effect \citep{Ragazzoni2020b}, the peak surface brightness of the trails on detector in equivalent $mag/arcsec^2$ is finally given by:
\[
m_{\rm{sur}} = m_{\rm{app}} - 2.5\lg(\frac{t_{\rm{eff}}}{t_{\rm{exp}}}) - 2.5\lg\frac{1 arcsec^2}{\pi(\theta_{\rm   sat}/2)^2}
\]

Note that the surface brightness derived here is the peak brightness rather than the average brightness of a trail. We need to have a proper view of what this surface brightness represents, otherwise there is a suspicion of overestimating the impact.

\subsection{Surface brightness thresholds}\label{subsec:thresholds}

By simulating the response of the LSST camera to the satellite streaks in a given scenario, \citet{Tyson2020a} (see Table 1) provides the satellite apparent magnitude and streak surface brightness corresponding to the saturation
level of 100,000$e^-$ in the center pixel, as well as the satellite apparent magnitude corresponding to the best-case crosstalk correctable limit of 10,000$e^-$. The streak surface brightness corresponding to the best-case crosstalk correctable limit can be derived by:
\[
m_{\rm tra}^{\rm X} =  m_{\rm sta}^{\rm X} - m_{\rm sta} + m_{\rm tra}
\]
where $m_{\rm sta}, m_{\rm sta}^{\rm X}$ represent the satellite stationary magnitudes for the saturation
level and the best-case crosstalk correctable limit, respectively. These surface brightness thresholds serve as the basis for identifying streak hazards. Note that each band in \citet{Tyson2020a} has a 30s exposure, and we correct these thresholds for bands other than the $u$ band to a 15s exposure by subtracting 0.75 mag. For instance, the apparent magnitude thresholds of 2.89 mag, 5.87 mag and a recommended dimming target of 7 mag for the g-band correspond to corrected surface brightness thresholds of approximately 14.91 mag, 17.89 mag and 19.02 mag, respectively. These surface brightness thresholds are used to assess the impacts quantitatively in Section \ref{sec:result}.

\section{results} \label{sec:result}
\subsection{satellite apparent magnitude above observatory}\label{subsec:app_mag}
\begin{figure}[ht!]
\epsscale{1.2}
\plotone{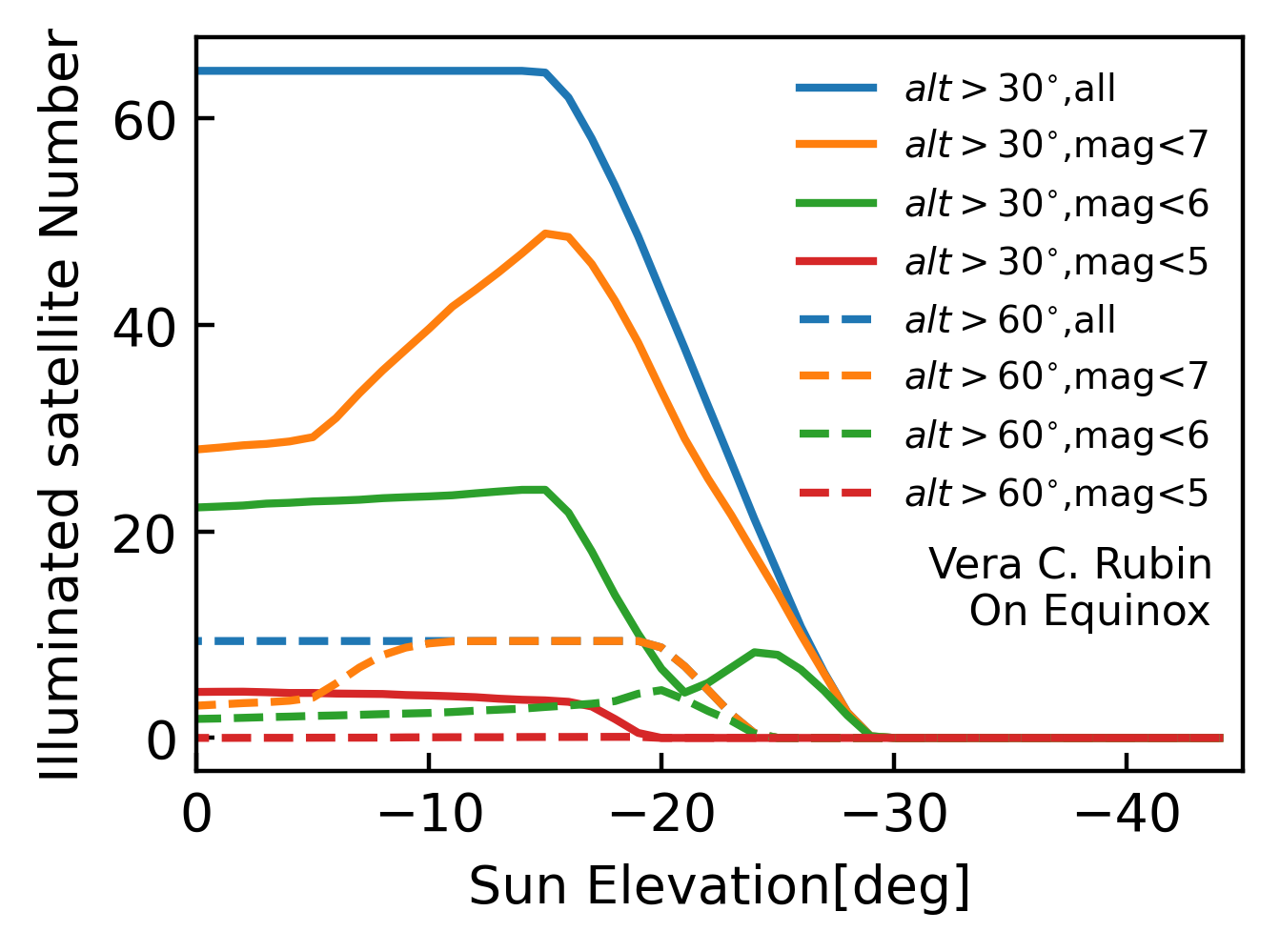}
\caption{Number of visible satellites at different solar altitudes. It is drawn for the Rubin Observatory before midnight on the spring equinox with totally 11,908 satellites on orbit. Despite differences in solar azimuth at the same solar altitude in other seasons or after midnight, variations in satellite numbers are statistically relatively small from simulations.\label{fig:app_mag}}
\end{figure}

To assess the impact of Starlink satellites on LSST, we start with an analysis of the number of illuminated satellites of different apparent magnitudes above the Rubin Observatory, as shown in Figure~\ref{fig:app_mag}.

According to the constellation outlined in Table ~\ref{tab:orbits}, there are a maximum of 64.60 illuminated satellites above $30^{\circ}$ elevation, with an average of $65.90\%$ of them brighter than 7 mag during twilight (defined as solar altitude between $-6^{\circ}\sim-18^{\circ}$) and $69.33\%$ during both twilight and nighttime. An average of 9.38 illuminated satellites will always be above $60^{\circ}$ elevation during twilight and all brighter than 7 mag.

When the solar altitude is reduced during twilight, satellites remain visible around zenith, so the number of visible satellites has not decreased and even increased significantly for those brighter than 7 mag. This is due to the increased brightness for satellites around the solar azimuth direction \citep{lu2024}. The number of illuminated satellites decreases rapidly into the nighttime, and no visible satellites above $30^{\circ}$ elevation when the solar altitude is below $-29^{\circ}$. Therefore, the impact from an apparent magnitude viewpoint is interestingly variable and can be simply summarised as increasing and then decreasing as the sun decreases.

\subsection{trail surface brightness across sky}\label{subsec:all_sky}

\begin{figure}[ht!]
\epsscale{1.2}
\plotone{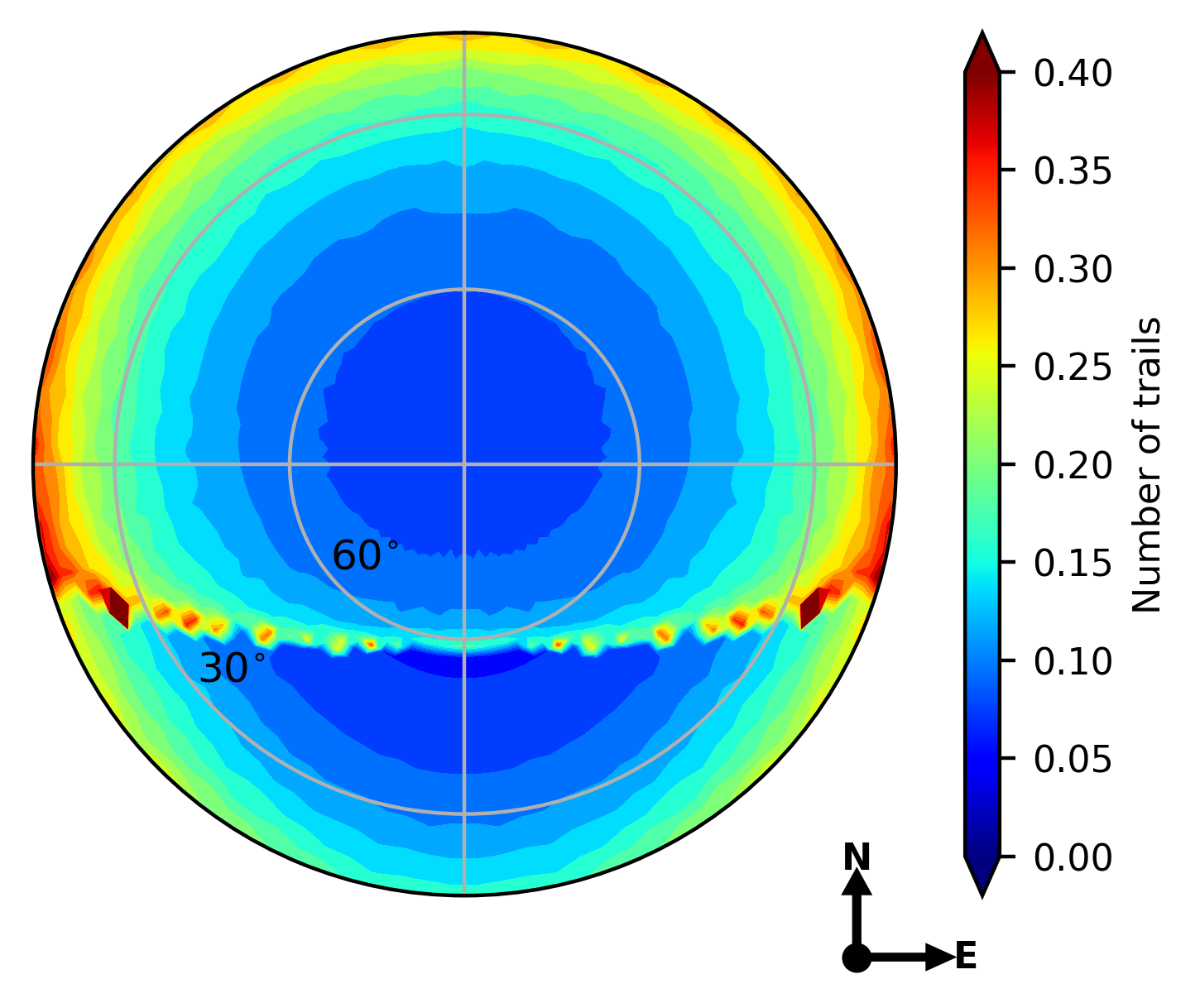}
\plotone{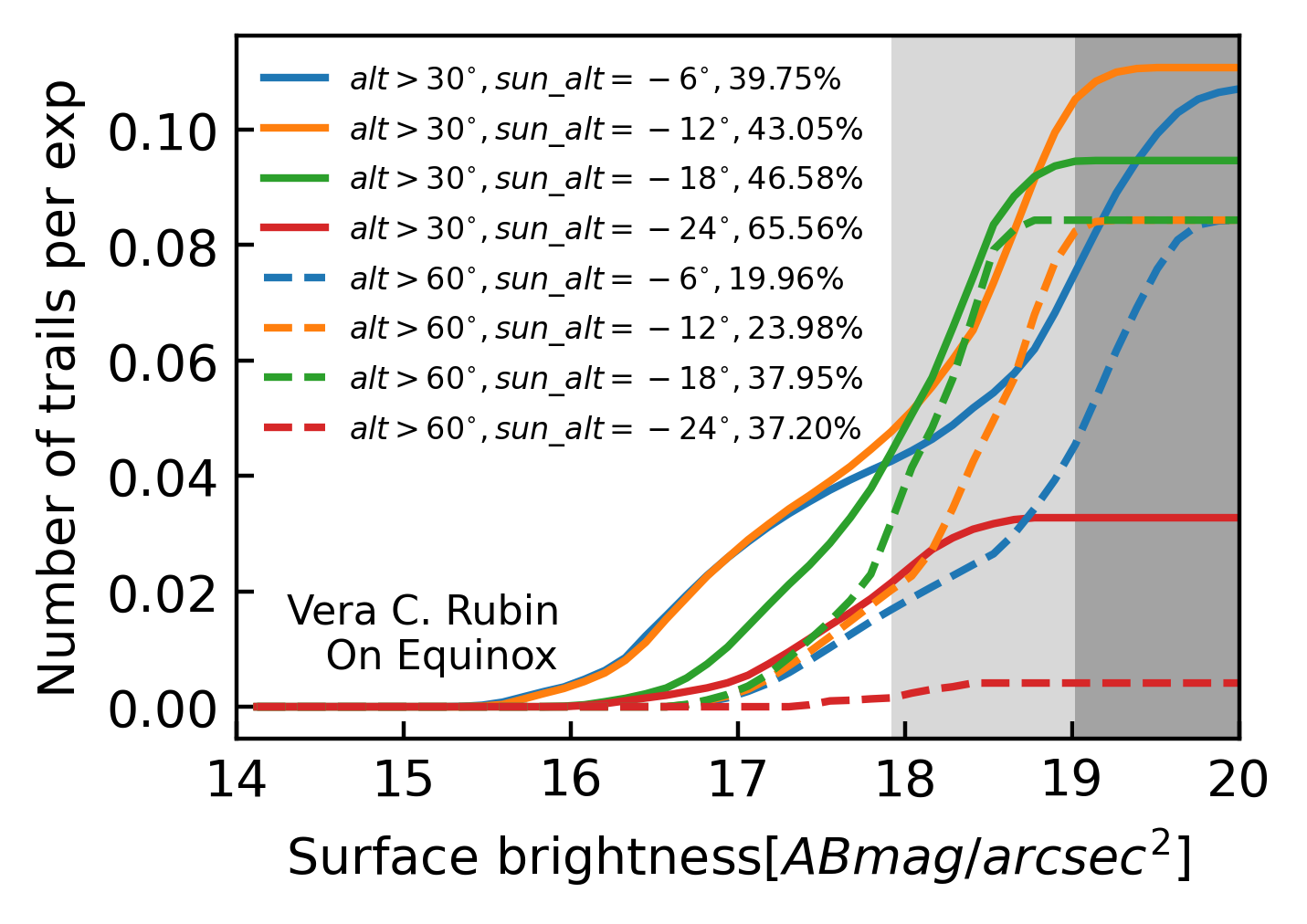}
\caption{Distribution of the number of Starlink trails in a LSST exposure (top). The Starlink constellation is assumed to contain 11,908 satellites. All satellites above are illuminated and the exposure is set to 15 seconds. Peak surface brightness distribution of satellite trails in all-sky simulations (bottom). The shaded areas represent the best-case crosstalk correctable limit and the recommended dimming target, respectively. No trails can reach the saturation-level magnitude, corresponding to a surface brightness of 14.91 mag in $g$ band. The percentage of streaks brighter than the best-case crosstalk correctable limit is listed in the legend. The worst case occurs when the solar altitude is $-15^{\circ}$ for observations above $30^{\circ}$ elevation. This is plotted on the spring equinox, with minimal variation for the other seasons.\label{fig:trail_num_and_sur_bri_hist_allsky}}
\end{figure}

Due to spatial variation in trail density, the likelihood of a particular bright trail appearing in an exposure is determined by multiplying the area of the corresponding sky region by its trail density. Figure~\ref{fig:trail_num_and_sur_bri_hist_allsky} (top) shows the number of trails for LSST in a 15s exposure with a solar altitude of $-6^{\circ}$, when most satellites over the station can be illuminated. As the sun descends, the satellites become invisible in some sky regions, but in others, the trail number remains almost constant. Overall, the larger the zenith distance, the greater the number of trails. That is why NEO searches will be most affected by satellites, which needs to observe at low elevation during twilight. The figure shows a band across the map with a significant excess of trails, which is caused by the shell with a $33^{\circ}$ inclination in Gen2A, and is visible to Rubin observatory (see \citet{Lawler2022b} for a related analysis).

The cumulative distribution of the number of streaks per exposure with surface brightness is also shown in Figure~\ref{fig:trail_num_and_sur_bri_hist_allsky} (bottom). For the constellation assumed in this letter, a streak occurs on average for every $9$ exposures above $30^{\circ}$ elevation during twilight for LSST, and about half of these trails are within correctable brightness. There are fewer trails and smaller proportion of bright ones above $60^{\circ}$ elevation. 

Comparison of the curves above $60^{\circ}$ in twilight (solar altitude of $-6^{\circ},-12^{\circ},-18^{\circ}$) clearly shows that although the total number of trails is the same (the curves end up aligned), the proportion of bright trails increases as the solar altitude descends, consistent with the trend of satellite apparent magnitude in Figure~\ref{fig:app_mag}. Therefore, when scheduling observations during twilight, the non-monotonic impact of satellite streaks on LSST as solar altitude decreases should be considered.

\begin{figure*}[ht!]
\plotone{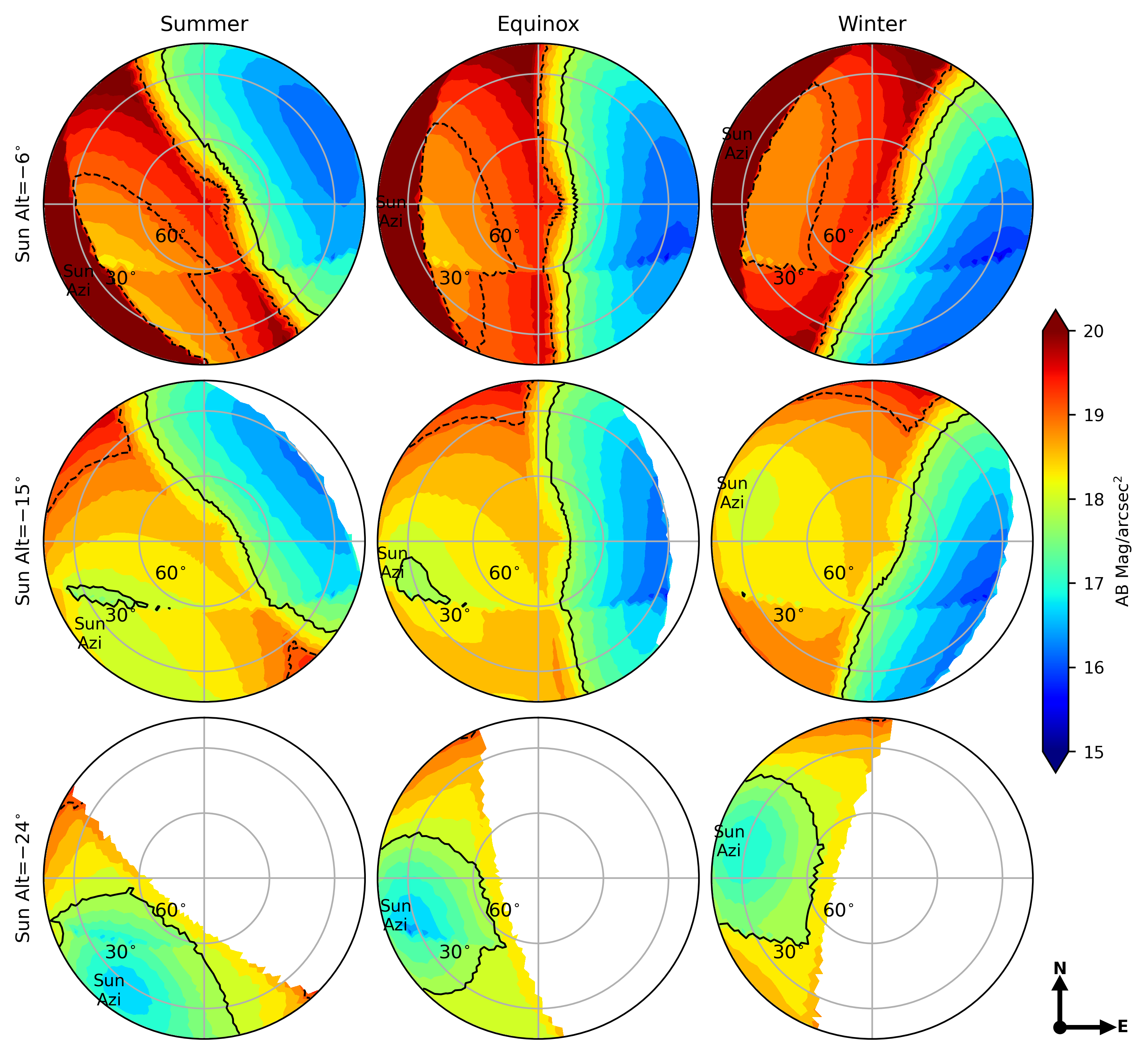}
\caption{All sky maps of g-band average peak surface brightness of satellite trails above Rubin observatory. The occurrence probability of satellite trails in different observational directions is used as the weights for averaging the trail surface brightness. The solid and dashed boundary lines represent the thresholds corresponding to the best-case crosstalk correctable magnitude and the recommended dimming target, respectively. A distribution feature related to the solar azimuth is clearly exhibited in the figure. \label{fig:sur_bri_map}}
\end{figure*}

The spatial distribution of the peak surface brightness over the station could reveal the impact of satellite trails on astronomical observations in various directions. However, the surface brightness of trails in an exposure may vary due to the different altitudes and inclinations of the satellites in different orbital shells. Further, satellites in the same shell are divided into ascending and descending cases, and although their apparent magnitudes are the same, their different apparent angular velocities lead to differences in surface brightness. As a result, there will be multiple satellites with different brightnesses and different occurrence probabilities in a FOV. By utilizing LSST parameters, we can acquire the g-band trail surface brightness for different pointings and plot its distribution across the sky (using AltAz coordinates) in Figure~\ref{fig:sur_bri_map}, where the value in a direction is averaged by weighting the likelihood of trail with different surface brightness.

Figure~\ref{fig:sur_bri_map} illustrates how the surface brightness distribution is affected by the position of the sun, i.e., the sky appears brighter around the anti-solar direction, while the opposite directions become gradually brighter as the solar altitude decreases. The distribution also varies with seasonal changes in solar azimuths.  

The solid and dashed boundary lines in Figure~\ref{fig:sur_bri_map} represent the thresholds corresponding for the best-case crosstalk correctable limit and the recommended dimming target respectively, as discussed in Section ~\ref{subsec:thresholds}. The figure reveals that a large portion of the sky exceeds the correctable limit, while areas fainter than the dimming target are only satisfied in some low altitude directions during twilight.

\subsection{trail surface brightness in schedule}\label{subsec:schedule}

To provide a more accurate evaluation, the peak surface brightnesses of trails appearing in the simulated LSST schedules are analyzed. These schedules, taken from \citet{Hu2022}, were created by adding a satellite constellation avoidance strategy to the baseline, but did not consider the varying brightness of the streaks. These schedules were simulated for several constellations respectively (Starlink Gen1, Gen2 and Oneweb) with different dodging weights. The schedules used to avoid Starlink Gen2 are adopted in the following analyses, as our constellation setup is relatively closer to that.

\begin{figure*}[ht!]
\epsscale{1.1}
\plottwo{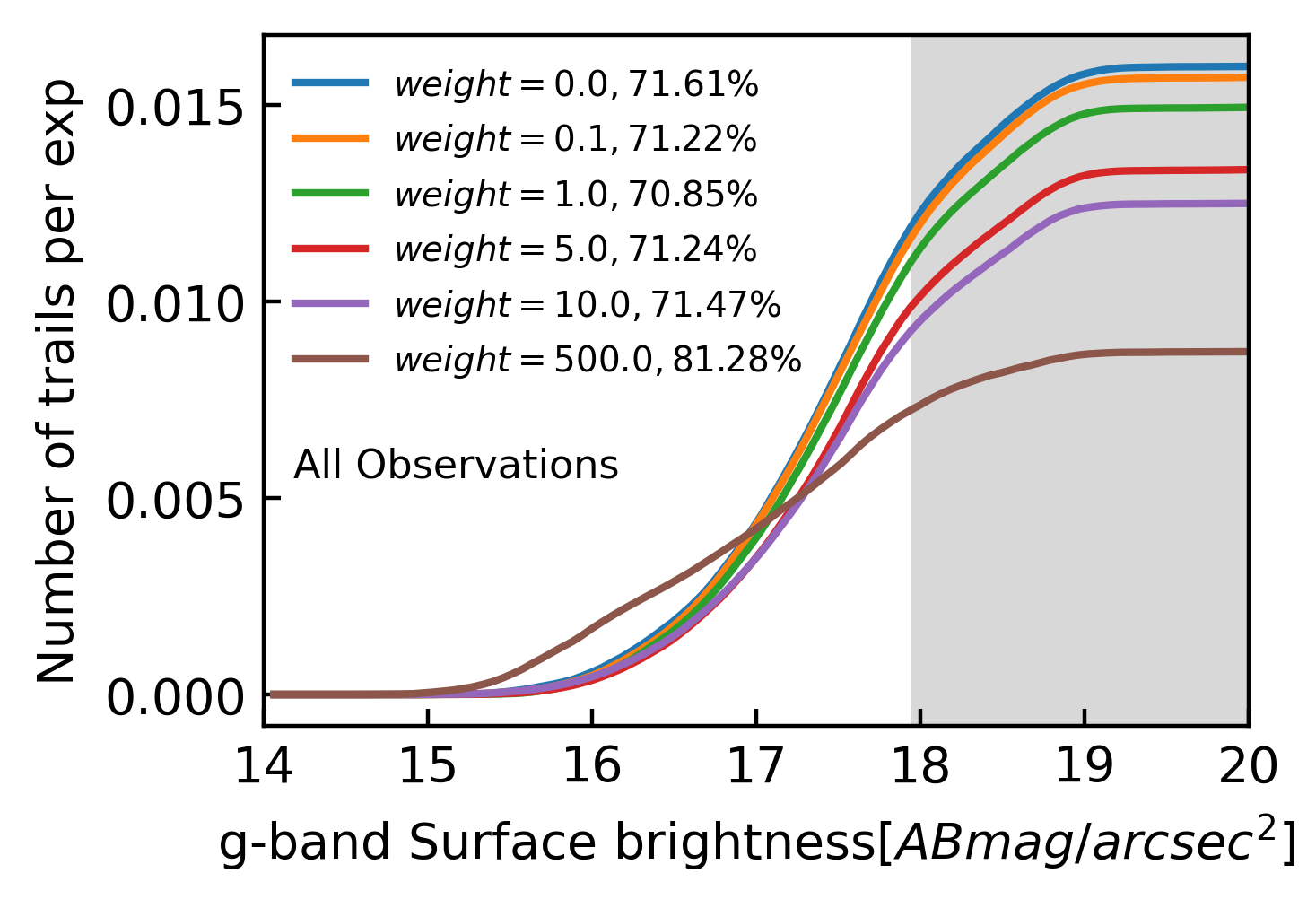}{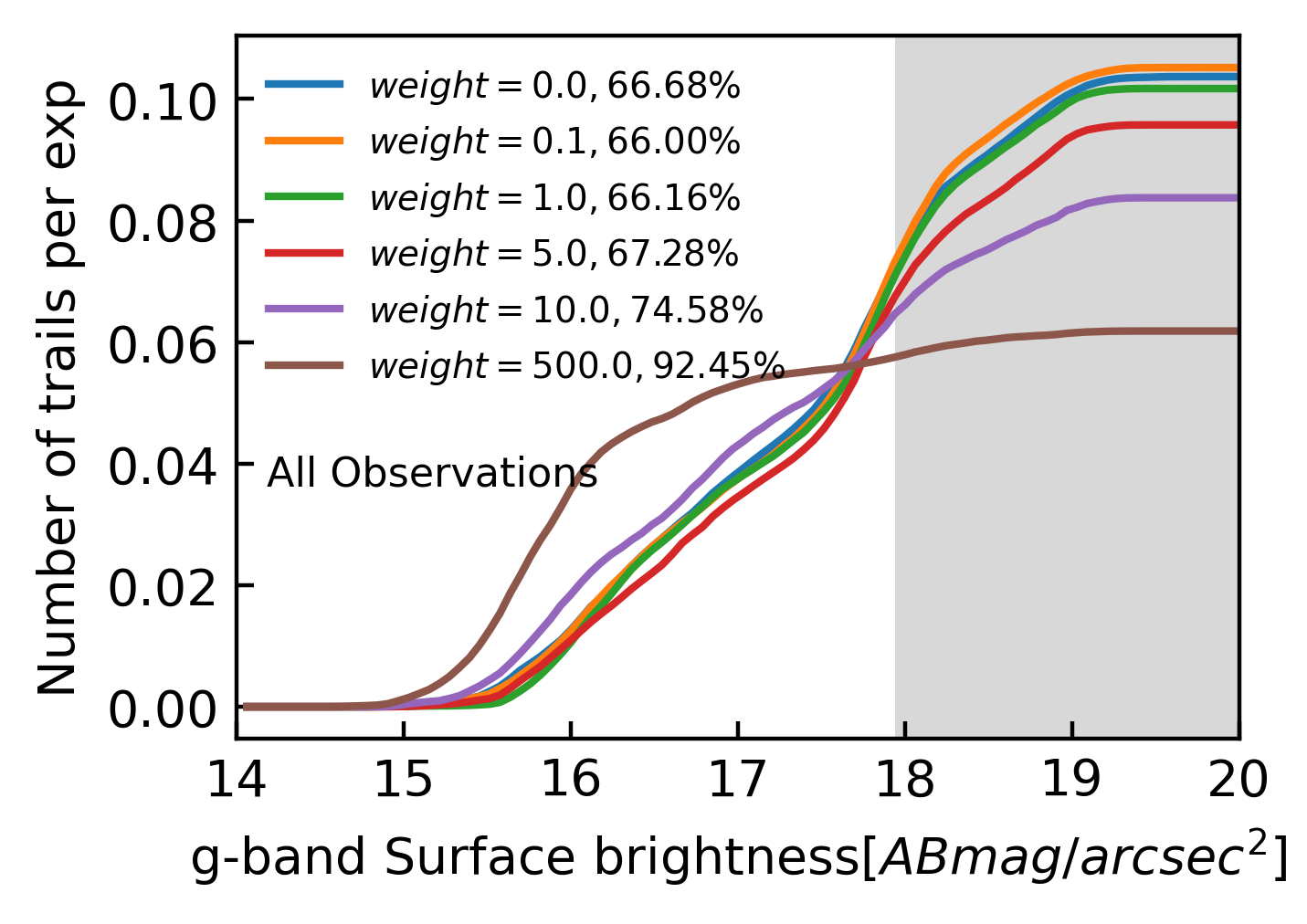}
\caption{Peak surface brightness distribution of satellite trails in simulated schedules. Consider full observations (left) and observations below $50^{\circ}$ elevation during twilight (right), separately. The shaded areas represent the crosstalk correctable limit, corresponding to a surface brightness of 17.89 mag in $g$ band. The percentage of bright streaks is listed in the legend. Trail brightness in other bands is adjusted to align their best-case crosstalk correctable limits with that of g-band. Thus the surface brightness of the figure x-axis is meaningless for other bands except the g band.\label{fig:sur_bri_schedule}}
\end{figure*}

To count the number of streaks per exposure within all filters together, we shift the brightness of all other bands so that their best-case crosstalk correctable limits align with the g-band limit of 17.89 $\rm{mag/srcsec^2}$. Figure~\ref{fig:sur_bri_schedule} left panel is the cumulative distribution of all exposures in schedules, indicating that satellite trails are encountered on average per 63 exposures without satellite avoidance. When considering only twilight and low elevations for typical NEO searches, Figure~\ref{fig:sur_bri_schedule} right panel shows that the impact is approximately seven times worse than that of the overall schedule. 

As the dodging weight increases, the chance of trails appearing is reduced by almost half at the maximum weights in both cases. However, a larger dodging weight would result in a higher percentage of bright trails, as shown in Figure~\ref{fig:sur_bri_schedule}, in addition to more wasted observing time, as noted in \citet{Hu2022}. To quantify the percentage of bright streaks in the schedules, only the best-case crosstalk correctable limits are used to determine the bright trails, as there are no saturation-level trails and very few are darker than the recommended magnitude. The percentage of bright streaks increases from $71.61\%$ at zero weight to $81.28\%$ at maximum weight in regular observations, and from $66.68\%$ to $92.45\%$ for observations in low elevations during twilight. The results confirm that satellite avoidance algorithms for the scheduling of projects like LSST must consider both the number and the surface brightness of trails. Otherwise, there may be unexpected situations where the number of streaks is reduced, but the streaks are brighter on average.

\section{discussion}
An important feature of our work is that it uses a relatively reliable near-term plans for the Stralink constellation with a total of 11908 satellites, to highlight the potential impact in the early phases of LSST operations. If future larger constellations have similar orbital altitudes and inclinations to the current, 
the impact can be estimated by a linear relationship with the number of satellites. Otherwise, it may be necessary to recalculate the chances of different brightness streaks appearing in an exposure based on new orbital parameters.

According to the photometric model of the Starlink v1.5 satellite used in this study, only a few trails are fainter than the recommended brightness. Future versions of the satellite should strive towards this goal with the collaboration of satellite operators and astronomers. The brightness of the new v2.0mini satellite is reported to have been significantly reduced \citep{mallama2023assessment} by several effective mitigation techniques, including the application of a new dielectric mirror film, an opaque pigment for solar backsheet, a low reflectivity black paint as well as a mitigation-oriented off-pointing maneuver by solar array when crossing the terminator \citep{SpaceX2022}. We will continue to focus on the impact of the new satellites on astronomical observations.

\begin{acknowledgments}
We would like to thank Elena Katkova for the efforts in maintaining the MMT satellite photometric database. This research was supported by the Natural Science Foundation of Jiangsu Province of China(No. BK20221164).
\end{acknowledgments}

\vspace{5mm}

\software{numpy \citep{harris2020array}, 
astropy \citep{price2018astropy}, 
skyfield \citep{rhodes2019skyfield}, 
healpix \citep{gorski2005healpix}, matplotlib \citep{hunter2007matplotlib}.
          }



\bibliography{sample631}{}
\bibliographystyle{aasjournal}
\end{document}